\documentclass[10pt,letterpaper,amsmath]{article}
\usepackage{opex3}

\begin{document}

\newcommand{\km} {\overline{k}}
\newcommand{\lm} {\overline{l}}
\newcommand{\crJI}{\mathrm{corr}[J,I]}
\newcommand{\cvJI}{\mathrm{covar}[J,I]}

\title{Simultaneous measurement of the microscopic dynamics and the mesoscopic displacement field in soft systems by speckle imaging}

\author{L. Cipelletti$^{1,2}$, G. Brambilla$^{1,2,3}$, S. Maccarrone$^{1,2}$, S. Caroff$^{1,2}$}

\address{$^1$Universit\'{e} Montpellier 2, Laboratoire Charles Coulomb UMR 5221, F-34095 Montpellier,
France\\
$^2$CNRS, Laboratoire Charles Coulomb UMR 5221, F-34095 Montpellier, France\\
$^3$ Present address: Formulaction, 31240 L'Union, France}

\email{luca.cipelletti@univ-montp2.fr} 



\begin{abstract}
The constituents of soft matter systems such as colloidal suspensions, emulsions, polymers, and biological tissues undergo microscopic random motion, due to thermal energy. They may also experience drift motion correlated over mesoscopic or macroscopic length scales, \textit{e.g.} in response to an internal or applied stress or during flow. We present a new method for measuring simultaneously both the microscopic motion and the mesoscopic or macroscopic drift. The method is based on the analysis of spatio-temporal cross-correlation functions of speckle patterns taken in an imaging configuration. The method is tested on a translating Brownian suspension and a sheared colloidal glass.
\end{abstract}




\section{Introduction}
Rough surfaces and scattering media generate a characteristic speckle pattern~\cite{Goodman2007} when illuminated by coherent light, \textit{e.g.} from a laser. By analyzing a time sequence of speckle patterns, valuable information can be retrieved on the sample evolution. Broadly speaking, one may distinguish between ``static'' speckle patterns generated by solid objects and ``dynamic'' speckles formed by soft matter systems (\textit{e.g.} colloidal suspensions, emulsions, polymer solutions, biological tissues), whose components undergo Brownian motion, thereby continuously reconfiguring the scattered speckle pattern. In the former case, relevant to metrology and interferometry~\cite{erf78}, a rigid displacement or a long wave-length deformation is often  measured, \textit{e.g.} in response to vibrations, applied load, or a change of temperature; speckle patterns are recorded onto a 2D detector such as a CCD or CMOS camera, using an imaging optics. By contrast, the microscopic (\textit{e.g.} Brownian) dynamics of soft systems is quantified by $g_2(\tau)-1$, the autocorrelation function of the temporal fluctuations of the scattered intensity. In these dynamic light scattering  measurements (DLS, a.k.a. photon correlation spectroscopy~\cite{Berne1976}) a point-like detector (\textit{e.g.} a phototube) is placed in the far field, where it collects light within a few speckles.

Recent developments have made the distinction between these two research fields increasingly fuzzy. On the one hand, imaging geometries similar to those for static speckles have been used to detect motion, \textit{e.g.} in vascular flow (\cite{briers01,draijer09} and references therein). Motion is typically quantified by inspecting the contrast of the speckle pattern, $<I^2>/<I>^2$, where $I$ is the local intensity and $\left < \cdot \cdot \cdot \right >$ is an average over a small region centered around the point of interest. The method is based on the fact that speckles are blurred and the contrast reduced in those regions where significant motion occurs~\cite{Bandyopadhyay2005}. Rigid motion has been measured also by laser speckle velocimetry~\cite{adrian84,dudderar88}, by applying spatial cross-correlation methods to speckle images, an approach similar to particle imaging velocimetry~\cite{RefPIV} and image correlation velocimetry~\cite{ICV} often used in fluid mechanics. On the other hand, CCD and CMOS cameras are now routinely used as detectors for DLS, especially for samples exhibiting slow dynamics, i.e. speckle fluctuations on time scales from a fraction of second up to several hours. In the first implementations of these so-called multispeckle approaches~\cite{WongRSI1993,KirschJChemPhys1996}, the far field detection scheme of traditional DLS was used and the intensity correlation function was calculated from
\begin{equation}
g_2(\tau)-1 = \left< \frac{\left<I_p(t)I_p(t+\tau)\right>_p}{\left<I_p(t)\right>_p\left<I_p(t+\tau)\right>_p}-1\right>_t \,,
\label{Eq:g2multispeckle}
\end{equation}
where $I_p(t)$ is the intensity of the $p\mathrm{-th}$ pixel at time $t$ and $< \cdot \cdot \cdot >_p$ and $< \cdot \cdot \cdot >_t$ indicate averages over pixels and time, respectively. In the far field detection scheme, each pixel receives light issued from the whole scattering volume and all pixels are associated to nearly the same scattering vector $q = 4\pi n \lambda^{-1}\sin \theta/2$, where $n$ is the solvent refractive index, $\lambda$ the in-vacuo laser wavelength and $\theta$ the scattering angle. As in traditional DLS, the decay time of $g_2(\tau)-1$ is related to the time it takes a scatterer to move (relative to the other scatterers) over a distance $\sim q^{-1}$ ~\cite{Berne1976}. A further step in bridging the gap between DLS and speckle imaging methods is represented by photon correlation imaging (PCI)~\cite{DuriPRL2009}. In PCI, DLS data are obtained by analyzing time series of speckle patterns acquired using a 2D detector and a low-magnification imaging optics. Similarly to conventional imaging, a given area of the detector corresponds to a well-defined region in the sample. Unlike conventional imaging, however, the image is formed using only light scattered in a narrow range of scattering vectors $q$ . This allows one to calculate a spatially-resolved version of Eq.~(\ref{Eq:g2multispeckle}), where pixel averages are performed on small regions of the detector, thereby providing information on the local dynamics. This method has been applied to systems whose microscopic dynamics are significantly heterogeneous in space, such as glassy or jammed soft matter~\cite{DuriPRL2009,maccarrone10}, and it has been extended to highly turbid media such as drying coatings~\cite{zakharov10} and granular systems~\cite{amon12}.

As highlighted by this short overview, previous works have focussed either on the measurement of the rigid displacement of a set of scatterers, regardless of any relative motion between them, or, conversely, on their relative motion due to the microscopic dynamics, regardless of any average drift component. Here, we present a new method that allows one to quantify in a single measurement the contribution of each of these phenomena to the evolution of speckle patterns formed in the imaging geometry~\cite{patent}. Figure~\ref{FIG:peak}, which will be discussed later, shows the essence of the method in a glimpse: the spatial cross-correlation of two speckle images taken at a time lag $\tau$ exhibits a peak, whose position and height yield the sample rigid shift and its internal dynamics over the time $\tau$, respectively. Applications of this method to gels submitted to gravitational~\cite{brambilla11} or internal~\cite{lieleg11} stress have been presented in previous publications, but the method itself was not discussed there. In this paper, we provide a detailed description of the algorithm used to implement the method, addressing in particular the challenges inherent to PCI experiments, i.e. the reduced size of the speckles and the need to process a very large number of images in a reasonably short time. Finally, we demonstrate our method by testing it on a model system, a suspension of Brownian particles contained in a cell displaced by a motor, and by measuring the velocity profile and the microscopic dynamics of a sheared colloidal glass.

\section{Sub-pixel Digital Imaging Correlation algorithm}

The first step in our method is to find the local rigid displacement with sub-pixel resolution. To this end, we use a cross-correlation technique inspired by
particle imaging velocimetry~\cite{RefPIV} and image correlation
velocimetry~\cite{ICV}. A time series of
images of the sample is taken, using a PCI setup. Each image is divided into a grid of regions of
interest (ROIs), corresponding to square regions of side $L$ in the
sample. Under a rigid drift, the speckle pattern in a ROI at time $t$ will appear in a shifted position in a
successive image taken at time $t+\tau$. If the scatterers undergo relative motion, in addition to a rigid shift, the shifted ROI will not be an identical copy of the original one. Still, the displacement field can be estimated by calculating by what
amount a ROI of the second image has to be back-shifted in order to
\textit{maximize} its resemblance with the corresponding ROI of the first
image. In practice, for  a given ROI the
shift along the horizontal and vertical directions, $\Delta x$ and
$\Delta y$, is determined by searching the maximum of
$\mathrm{corr}[I,J]$, the spatial crosscorrelation of the intensity
of the two images, defined by
\begin{equation}
\mathrm{corr}[J,I](k,l) =
\frac{\mathrm{covar}[J,I](k,l)}{\sqrt{\mathrm{var}[J]
\mathrm{var}[I] }} \label{Eq:corr}
\end{equation}
with
\begin{eqnarray}
\mathrm{covar}[J,I](k,l) = N^{-1}\sum_{r,c}J_{r,c}I_{r+k,c+l} -
 N^{-2}\sum_{r,c}J_{r,c} \sum_{r,c}I_{r+k,c+l} \label{Eq:covar}\\
\mathrm{var}[I] = N^{-1}\sum_{r,c}I_{r,c}^2 - \left (
N^{-1}\sum_{r,c}I_{r,c} \right ) ^2 \, .
 \label{Eq:var}
\end{eqnarray}
In the above equations, $I_{r,c}$ is the intensity at time $t$ of the pixel at row $r$
and column $c$, $J_{r,c}$ is the intensity at the same location but
at time $t+\tau$, and $k$ and $l$ are the shifts expressed in number
of rows and columns, respectively. Here and in the following, double sums over $r$ and $c$ extend over all rows and columns for which the terms of the sum exist, in this case the
$N$ pixels of the overlap region between the full image and the
shifted ROI. For computational efficiency, $\cvJI$ is usually
calculated in Fourier space~\cite{NumericalRecipes}. Note that $\cvJI \rightarrow 0$ far from
the peak, where $I$ and $J$ are uncorrelated.

The position $(\km, \lm)$ of the global maximum of $\crJI$ yields the desired
displacement along the direction of columns ($x$ axis) and rows ($y$ axis), $\Delta x =
\lm$ and $\Delta y = \km$ respectively, with pixel resolution. Several schemes have been proposed to improve this
resolution, \textit{e.g.} by calculating the centroid of
$\crJI$, or by fitting its peak to a 2-dimensional analytical function
such as a Gaussian. While both methods work well for broad, circularly symmetric peaks, they tend to be less robust when the peak is sharp or it has an asymmetric shape. The shape of the peak
is determined by the spatial autocorrelation of the intensity
pattern; for our speckle images, it depends on the shape and size of
the speckles, which may not be symmetrical, depending on the shape of the illuminated sample volume and the imaging optics~\cite{Goodman2007}. Moreover, the peak usually extends over just a few pixels, because one minimizes the speckle size in order to maximize the information content in the image. To overcome the limitations inherent to peak-based schemes, we use an alternative approach based on a least-square method that allows us to obtain the displacement field with a
typical resolution of a few hundredths of a pixel, with no
requirements on the shape or broadness of the peak and without using
any fitting function.

\begin{figure}
\includegraphics[width=13.cm]{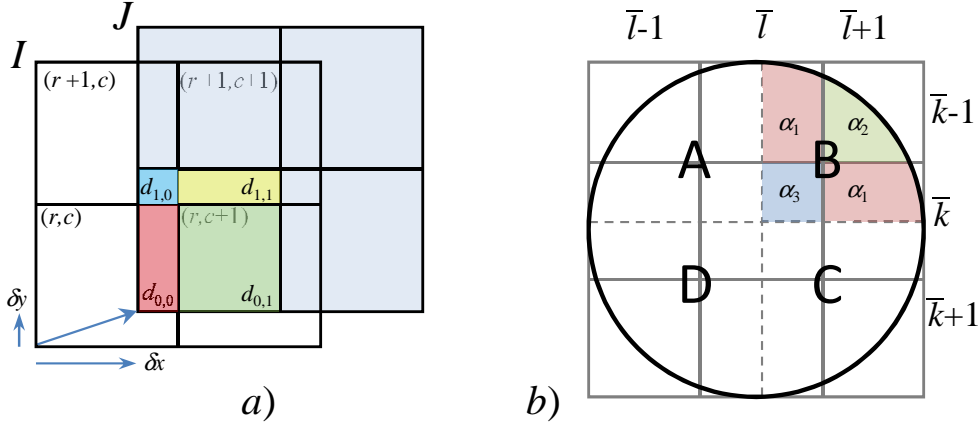}
\caption{a): schematic representation of an image $J$ that is a shifted version of image $I$ (only four pixels are shown for clarity). The intensity in a given pixel of $J$ may be obtained as a linear combination of (up to) four pixels of $I$, with weights proportional to the colored areas. b): quadrant detection scheme for locating the direction of the shift. The nine elements closest to the peak of the cross-correlation between $I$ and $J$ are represented here. See Sec.~\ref{sec:centerofmass} for more details.} \label{FIG:imageshift}
\end{figure}

The first step of our shift-finding algorithm is the same as in standard PIV
methods: $\crJI$ is calculated and the pixel-resolved shift, $(\km,\lm)$, is determined from the position of its global maximum. The next step consists in the refinement of such displacement
with sub-pixel resolution.
For the sake of simplicity, let us first assume that $J$ is simply a shifted version of $I$,
with displacement $(\Delta x = \lm+\delta x,\Delta y =
\km+\delta y)$, with $|\delta x| < 1$, $|\delta y| < 1$. The
intensity $J_{r,c}$ may then be expressed as a weighted average over
a suitable set of pixels of the intensity of the image $I$. In
principle, an infinite number of terms are needed to obtain
$J_{r,c}$, if $\delta x$ and $\delta y$ are
non-integer~\cite{RefTranslatingImages}. In practice, linear
interpolation is usually sufficient to reconstruct $J$ to a good
approximation, thereby greatly simplifying the calculation. Using
linear interpolation, the intensity $J_{r,c}$ at a given pixel is
expressed as the weighted sum over the (at most) four pixels of $I$ that
partially overlap with that pixel, as exemplified in Fig.~\ref{FIG:imageshift}a. Thus,
\begin{equation}
J_{r,c} = d_{0,0}I_{r+\km,c+\lm} +
d_{1,0}I_{r+\km+1,c+\lm}+d_{0,1}I_{r+\km,c+\lm+1}+d_{1,1}I_{r+\km+1,c+\lm+1} + \epsilon_{r,c} \,.
\label{Eq:linearcomb}
\end{equation}
The coefficients $d$ are the overlap areas shaded in
Fig.~\ref{FIG:imageshift}a: $d_{0,0} = (1-\delta x) (1 - \delta y)$,
$d_{1,0} = \delta x(1 - \delta y)$ and similarly for the other
terms. The term $\epsilon_{r,c}$ has been added to account for the fact that in general $J$ will not be an exact (albeit shifted) replica of $I$, because of experimental noise and due to any evolution of the speckle pattern, which in our case is due to the microscopic dynamics of the scatterers. For distinct speckles, these fluctuations are uncorrelated~\cite{Goodman2007,Berne1976}; we thus treat $\epsilon$ as a noise term and determine the displacement $(\Delta
x,\Delta y)$ as the rigid shift that minimizes, in a least-squares sense, the difference
between $J$ and the linear combination of $I$ in the r.h.s. of Eq.(\ref{Eq:linearcomb}). More specifically, we search for a set of four coefficients $\mathbf{a} = \{a_1,a_2,a_3,a_4\}$ that minimizes the cost function $\chi^2$ defined as
\begin{eqnarray}
\chi^2(\mathbf{a}) = \sum_{r,c} \sum_{i=1}^{4} \left
(a_{i}I_{r+k_i,c+l_i}-J_{r,c} \right )^2 \,,\label{Eq:chi22}
\end{eqnarray}
where explicit expressions relating $\mathbf{a}$ to the sub-pixel shift and $k_i$, $l_i$ to the pixel-resolved shift will be provided in the following. Note that in principle nine coefficients are required, instead of the four introduced here, since the direction of the shift is not known \textit{a priori}. However, we expect that only up to four of
them differ significantly from zero. In order to speed
up the determination of $\mathbf{a}$, we calculate the centroid of the peak of $\crJI$ in order to
predetermine the direction of the shift (\textit{e.g.} top-left, top-right
etc.), so that only four coefficients have to be computed. This is done by calculating in which of the four quadrants labeled by $\mathrm{A},...,\mathrm{D}$ in Fig.~\ref{FIG:imageshift}b lays the center of mass of the crosscorrelation peak, as explained in detail in Sec.~\ref{sec:centerofmass}, where we also provide explicit expressions for $k_i, l_i$. Note that the centroid algorithm is only used to determine the direction of the shift, not its subpixel magnitude, thus avoiding the limitations recalled above for sharp or non-symmetric peaks.

The optimum shift is obtained by imposing $\partial \chi^2 /
\partial a_{i} = 0$, $i=1,..,4$. By exchanging the order of the sums in Eq.~(\ref{Eq:chi22}) one recognizes that $\mathbf{a}$ is the solution of $\mathbf{b} = \mathbf{M}\cdot \mathbf{a} $,
with:
\begin{eqnarray}
b_i = \cvJI(k_i,l_i) \label{Eq:b}\\
M_{i,j} = \mathrm{covar}[I,I](k_i-k_j,l_i-l_j)\,. \label{Eq:M}
\end{eqnarray}
Any standard method is suitable to solve the above set of equations; in our implementation, we use singular value
decomposition~\cite{NumericalRecipes}. It should be emphasized that
$\mathbf{b}$ is known, since $\cvJI$ has been already calculated to estimate the pixel-resolved displacement (see Eqs.~(\ref{Eq:corr},\ref{Eq:covar})). Therefore, the only extracost
required for calculating the displacement with sub-pixel resolution
is the computation of $\mathrm{covar}[I,I]$ and the solution of the
set of linear equations $\mathbf{b} = \mathbf{M}\cdot \mathbf{a} $. This moderate
computational extracost is due to the fact that a linear
interpolation scheme has been adopted in Eq.~(\ref{Eq:chi22}):
higher-order interpolations, although more precise, would lead to a
much more complex, non-linear minimization problem. Finally, we note
that in a typical multispeckle DLS experiment, one calculates the correlation
functions for a given starting time (i.e. a given image $I$) and
several time delays $\tau$ (i.e. several images $J$). Therefore, the
computational cost for calculating $\mathrm{covar}[I,I]$ is shared
between several lags, further increasing the efficiency of the
algorithm.

Once $\mathbf{a}$ is computed, the shifts along the $x$ and $y$
directions are calculated with subpixel precision according to
\begin{eqnarray}
\Delta x = \frac{a_2+a_4}{\sum_{i=1}^{4}a_i} + l_1 \\
\Delta y = \frac{a_3+a_4}{\sum_{i=1}^{4}a_i} + k_1 \,
\end{eqnarray}
(see Eqs.~(\ref{Eq:k1}) and (\ref{Eq:l1}) in Sec.~\ref{sec:centerofmass} for the definition of $k_1$ and $l_1$).

\section{Dynamic Light Scattering: corrections to $g_2(\tau)-1$ for drifting samples}
In order to quantify the internal dynamics, one needs to compute the intensity correlation function $g_2-1$ between image $J$ and a shifted version of $I$, so as to avoid any artifact due to the rigid shift of the speckles.
Denoting by $I'$ the image $I$ shifted by $(\Delta x, \Delta y)$,
the (un-normalized) intensity correlation function corrected for the shift
contribution is
\begin{equation}
G_2(\tau) = N^{-1}\sum_{r,c}J_{r,c}I'_{r,c} \,. \label{eq:G2shift}
\end{equation}
The shifted image $I'$ may be constructed using an interpolation method. Tests on real speckle images show that linear interpolation, although suitable for determining the shift with good accuracy, is not precise enough to reconstruct a shifted version of $I$ suitable for the calculation of $g_2-1$. Higher-order interpolation schemes are thus required.
As shown in Ref.~\cite{RefTranslatingImages}, image shifting by interpolation is equivalent to
convolving the original image with a suitable kernel:
\begin{equation}
I'_{r,c} = \sum_{k,l}h(r+\Delta y-k)h(c+\Delta x-l)I_{k,l} \,,
\label{eq:shift_interpolation}
\end{equation}
where we have assumed for simplicity that the kernel is symmetrical
and separable, i.e. that $_{2D}h(x,y) = h(x)h(y)$. Unfortunately, in our case this approach would be too time-consuming, because it requires a convolution operation, Eq.~(\ref{eq:shift_interpolation}), in addition to the calculation
of the correlation function, Eq.~(\ref{eq:G2shift}).

We introduce below an alternative method that leads to a much faster algorithm,
where the only computational cost is that of evaluating the kernel
for a few points, with no need for the calculation of convolution
and correlation functions. It is convenient to consider only non-negative fractional shifts $\widetilde{\delta x}$, $\widetilde{\delta y}$, given by
\begin{eqnarray}
\Delta x = j_x + \widetilde{\delta x} \label{eq:delta1}\\
\Delta y = i_y + \widetilde{\delta y} \,, \label{eq:delta2}
\end{eqnarray}
with $j_x = \mathrm{floor}(\Delta x)$, $i_y = \mathrm{floor}(\Delta y) $,
where $\mathrm{floor}(x)$ is the largest integer $\le x$.  As we shall discuss it below, the
choice of the kernel is not crucial; a good choice is a truncated, windowed
$\mathrm{sinc}$ function with an even number, $M$, of supporting
points:
\begin{eqnarray}
h(x) &=  w(x)\sin(\pi x)/(\pi x)~~&\mathrm{for~}|x| \le M/2\\
h(x) &= 0 &\mathrm{elsewhere}\nonumber\,,
\end{eqnarray}
where we choose the three-term Blackman-Harris window function~\cite{RefTranslatingImages,harris78} defined as
\begin{equation}
w(x) = 0.42323+0.49755\cos\left(\frac{2\pi x}{M}\right)+0.07922\cos\left(\frac{4\pi x}{M}\right)\,.
\end{equation}
With this choice, the kernel is DC-constant~\cite{RefTranslatingImages,harris78}, i.e.
\begin{equation}
\sum_{k,l=-M/2+1}^{M/2}h(x) = 1\,,
\label{eq:dc_constant}
\end{equation}
a property that will be of use in the following.

Using Eqs.~(\ref{eq:delta1},\ref{eq:delta2}), the convolution
product (\ref{eq:shift_interpolation}) may be rewritten as
\begin{equation}
I'_{r,c} = \sum_{k,l=-M/2+1}^{M/2}h(\widetilde{\delta
y}-k)h(\widetilde{\delta x}-l)I_{k+r+i_y,l+c+j_x} \,.
\label{eq:convolution}
\end{equation}
By replacing the r.h.s. of Eq.~(\ref{eq:convolution}) in
Eq.~(\ref{eq:G2shift}) and by exchanging the order of the sums, we
obtain:
\begin{equation}
G_2(\tau) = \sum_{k,l=-M/2+1}^{M/2}h(\widetilde{\delta
y}-k)h(\widetilde{\delta x}-l) \left [N^{-1}\sum_{r,c}J_{r,c}I_{r+k
+i_y,c+l+j_x}\right] \,. \label{eq:G2shiftsmart}
\end{equation}
Finally, by recalling the definition of the covariance between $J$
and $I$, Eq.~(\ref{Eq:covar}), and using the fact that the kernel is DC-constant, Eq.~(\ref{eq:dc_constant}), one obtains
\begin{equation}
G_2(t,\tau) - \overline{J} \,\overline{I}= \sum_{k,l=-M/2+1}^{M/2}h(\widetilde{\delta
y}-k)h(\widetilde{\delta x}-l) \cvJI(k +i_y,l+j_x)
\end{equation}
or, equivalently,
\begin{equation}
g_2(t,\tau)-1 = \frac{\sum_{k,l=-M/2+1}^{M/2}h(\widetilde{\delta
y}-k)h(\widetilde{\delta x}-l) \cvJI(k +i_y,l+j_x)} {\overline{J}\, \overline{I}} \,,
\label{eq:heaven}
\end{equation}
where $\overline{I} = N^{-1}\sum_{r,c}I_{r,c} $ and similarly for $\overline{J}$.

Equation~(\ref{eq:heaven}) is the central result of our method. It
shows that the intensity correlation function corrected for the
shift contribution can be simply obtained as a linear combination of
a few terms of $\cvJI$, weighted by the kernel. Since $\cvJI$ has
already been calculated to determine the shift, the extra cost is
essentially limited to the evaluation of $M^2$ values of the kernel,
which is typically negligible. Finally, we note that $\cvJI$
vanishes on the length scale of the speckle size as its argument
departs from $(i_y,j_x)$, which is close to the location of the peak
of the covariance. Hence, it is sufficient to take $M$ on the order
of a few speckle sizes (in units of pixels), because in
Eq.~(\ref{eq:heaven}) the contribution of the kernel for larger lags
would be multiplied by a vanishingly small quantity. For example, we
find that for images with a speckle size of about 5 pixels, the
correction is virtually independent of $M$ for $M \ge 8$.

\section{Experimental tests}

We test our method on two samples: a suspension of Brownian particles loaded in a cell displaced by a motor, and a colloidal glass to which a shear deformation is applied. The Brownian sample is a dispersion of polystyrene microspheres (radius $R = 0.265~\mu\mathrm{m}$) in an aqueous solution of fructose at $75\%$ weight fraction. The particle volume fraction is $10^{-5}$ and the sample is kept at a temperature $T = 9~^{\circ}\mathrm{C}$. The setup is described in~\cite{ElMasriJPCM2005}; we use the imaging geometry shown in Fig. SM1 a) of~\cite{DuriPRL2009}, where an image of the sample is formed onto a CCD detector using light scattered at $\theta = 90~\mathrm{deg}$, corresponding to a scattering vector $q = 2.46~\mu\mathrm{m}^{-1}$. The field of view is $1820 \times 364 \mu\mathrm{m}^2$ and images are acquired at a rate of 10 Hz. The sample cell is attached to a motor that can impose a drift in the $y$ direction at a controlled speed, $v_y = 10 \mu\mathrm{m}~\mathrm{s}^{-1}$.

\begin{figure}
\includegraphics[width=13.cm]{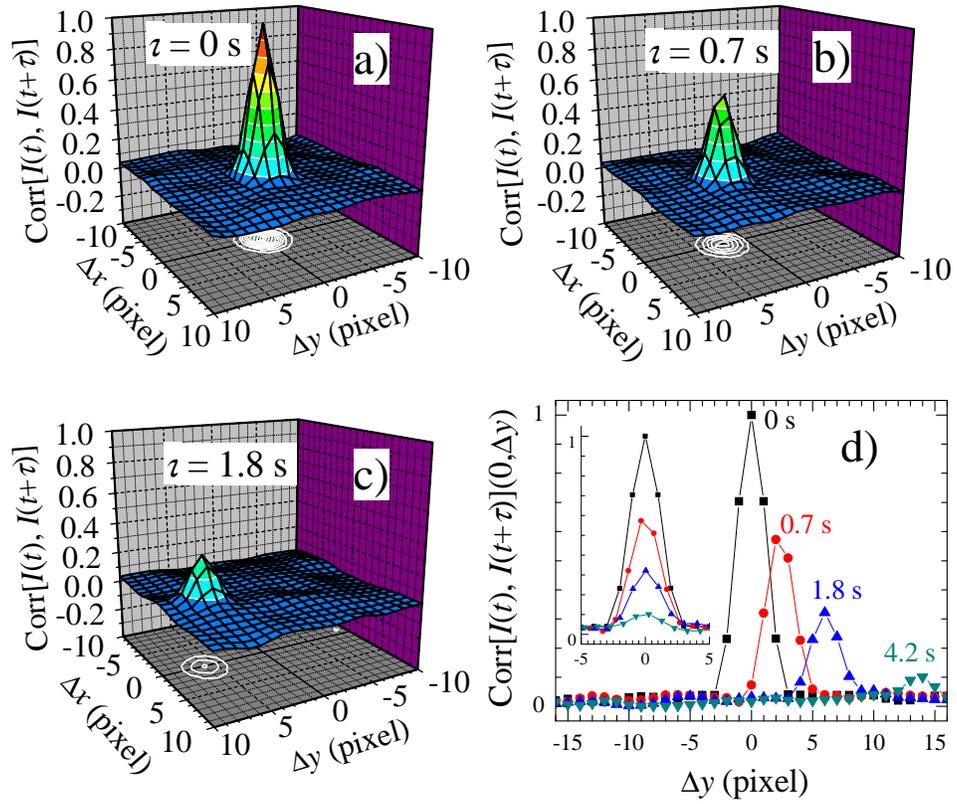}
\caption{a)-c) spatial cross-correlation between speckle images generated by a diluted Brownian suspension that is translated along the $y$ direction by a motor. The speckle patterns are recorded on a CCD using an imaging collection optics (see text for more details). As the delay $\tau$ between pair of images is increased, the peak position shifts to larger $y$ and its height decreases, due to the relative motion of the Brownian particles. d) cut of the cross-correlation along the $\Delta x =0$ line. Curves are labeled by $\tau$. Inset: same data, replotted as a function of spatial shift with respect to the peak position.} \label{FIG:peak}
\end{figure}

Figures~\ref{FIG:peak} a)-c) show the spatial crosscorrelation calculated applying Eq.~(\ref{Eq:corr}) to pairs of speckle images taken while displacing the sample, for three different time lags. For $\tau=0$ s, Eq.~(\ref{Eq:corr}) yields the spatial autocorrelation of the speckle pattern: accordingly, a sharp peak of height one and centered at $\Delta x = \Delta y = 0$ is visible, whose FWHM $\approx 2.9$ pixels provides the speckle size. As the lag increases, the peak position drifts in the $y$ direction, due to the translation motion imposed by the motor. Additionally, its height decreases, due to the relative motion of the Brownian particles that reconfigure the speckle pattern. Figure~\ref{FIG:peak} d) shows a cut of the crosscorrelation peak along the $\Delta x=0$ direction, for four values of $\tau$. From this plot, it is clear that if $g_2(\tau)-1$ was to be computed from a purely temporal crosscorrelation, as in Eq.~\ref{Eq:g2multispeckle}, one would observe a spurious, fast decay, essentially due to the rigid shift only. This would correspond to follow $\mathrm{corr}[I(t),I(t+\tau)]$ at $\Delta x = \Delta y = 0$, as a function of $\tau$. By contrast, if the relative motion of the Brownian particles is to be obtained, one has to measure the height of the peak as it drifts, as in the method proposed here. The inset of Fig.~\ref{FIG:peak} d) shows the same data, plotted as a function of distance along $y$ with respect to the (subpixel) peak position. It is worth noting that the peak width remains constant, in contrast to what suggested (but not demonstrated, to our knowledge) in patent literature~\cite{patentUSA}, where it was proposed that the peak would broaden with $\tau$ as a result of the internal motion of the scatterers. Thus, the relevant parameter for extracting the relative motion is indeed the peak height, not its width.

\begin{figure}
\includegraphics[width=13.cm]{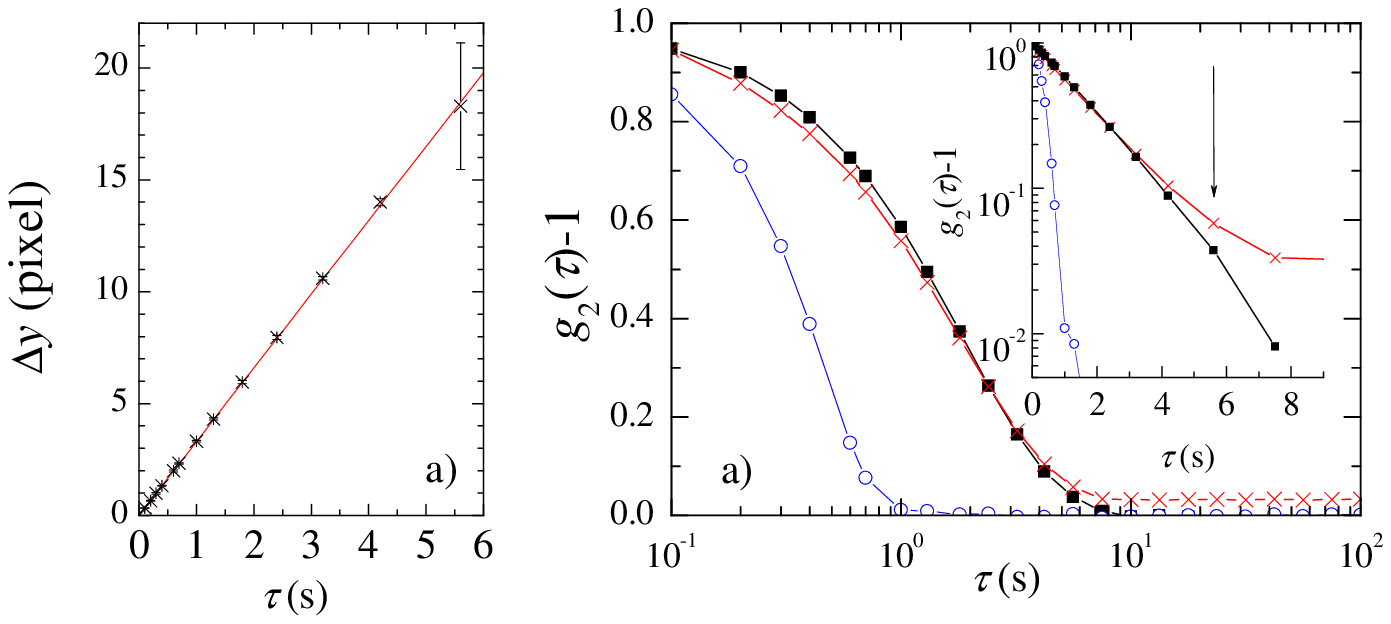}
\caption{a): displacement versus time, measured for a diluted Brownian suspension translated at a constant speed. The line is a linear fit to the data. b) Intensity correlation functions probing the microscopic dynamics. Solid squares: quiescent sample; open circles: raw $g_2-1$ measured while translating the sample with a motor; crosses: same data, corrected for the contribution of the rigid motion of the speckle pattern.} \label{FIG:brownian}
\end{figure}

In figure~\ref{FIG:brownian} a), we show the displacement of the speckle pattern as a function of $\tau$, obtained from the sub-pixel peak position averaged over 200 pairs of images (\textit{i.e.} 20 s), taken while translating the sample. The data are very well fitted by a linear law (red line) as expected for motion at constant speed, thus indicating that our algorithm captures very well the drift component of the speckle pattern, from a fraction of a pixel up to tens of pixels. The error bars are the standard deviation of the displacement over the measurement time, $\sigma_y$. For $\tau \le 4.2~\mathrm{s}$, $ \sigma_y/\Delta y < 4\%$ and the error bars are smaller than the symbol size, indicating that the detection of the peak position is very reliable, even when the peak height is as low as 0.1 or the displacement is just a fraction of a pixel. For $\tau = 5.6~\mathrm{s}$, the error bar is significantly larger, because the peak height becomes comparable to the noise level and the peak position can be hardly resolved. Beyond  $\tau = 5.6~\mathrm{s}$, no peak can be reliably found, thus preventing the displacement to be measured. The slope of the linear fit to the data is $3.30 \pm 0.03~ \mathrm{ pixel~s}^{-1}$. Recalling that the nominal speed of the motor is $v_y = 10 \mu\mathrm{m}~\mathrm{s}^{-1}$, this implies that 1 pixel corresponds to $3.03 \pm 0.03~\mu\mathrm{m}$ in the sample, in excellent agreement with $3.15 \pm 0.15~\mu\mathrm{m/pixel}$ as obtained from the magnification of the imaging system, evaluated using geometrical optics. Figure~\ref{FIG:brownian} b) shows the intensity correlation function $g_2(\tau)-1$, averaged over 20 s. If the sample is kept at rest during the measurement (black squares), the intensity correlation function exhibits an exponential decay, as expected for diluted Brownian suspensions~\cite{Berne1976}, as better seen in the inset that shows the same data in a semilog plot. When the sample is translated at constant speed, the uncorrected $g_2-1$ decays on a much shorter time scale (blue circles) and its shape departs from a simple exponential. Clearly, no information on the microscopic dynamics can be obtained from the uncorrected data. The red crosses are the data corrected according to Eq.~(\ref{eq:heaven}): for $\tau \le 4.2~\mathrm{s}$ the corrected $g_2-1$ is very close to that measured for the stationary sample, thereby demonstrating the effectiveness of our correction scheme. For larger lags, the corrected data tend to overestimate $g_2-1$: this is consistent with the fact that the displacement cannot be reliably measured, as discussed in relation to fig.~\ref{FIG:brownian} a). Indeed, in this case the peak-finding algorithm spuriously interprets the highest level in the noisy base line of $\mathrm{corr}[I(t),I(t+\tau)]$ as the (higher-than-expected) degree of correlation.

\begin{figure}
\includegraphics[width=13.cm]{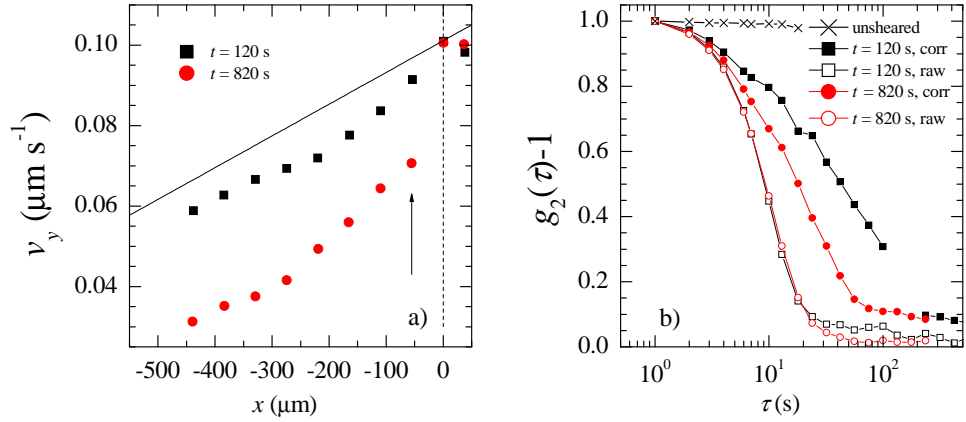}
\caption{a): velocity profiles for a sheared colloidal glass (symbols). Data are labeled by the time after initiating the shear. The dotted line indicates the position of the mobile wall, the solid line is the the velocity profile for uniform shear. The arrow indicates the location for which the data shown in b) have been measured. b) intensity correlation functions for the quiescent glass (crosses) and after applying a constant shear. Open (solid) symbols indicate the raw (corrected for drift) intensity correlation functions.} \label{FIG:sheared}
\end{figure}

Having validated our method on a model sample whose displacement is well controlled, we test it on a more realistic experimental situation, a sheared colloidal glass for which the velocity field is not uniform over the field of view. The sample is a dense suspension of hard-sphere-like colloidal particles, a widely-studied model system for the glass transition~\cite{PuseyPRL1987}. The particles have radius $\approx 100~\mathrm{nm}$ (as in~\cite{BrambillaPRL2009}) and volume fraction $\varphi \approx 0.6$. The setup is similar to that for the Brownian sample, but here the sample is kept in a square cell of section $10 \times 10~\mathrm{mm}^2$, in which a glass bead of diameter $D = 5~\mathrm{mm}$ is inserted. The bead is attached to a motor that displaces it in the $y$ (vertical) direction, parallel to the cell wall, at a speed $v_y = 0.1 ~\mu\mathrm{m~s}^{-1}$. The minimum gap $e$ between the wall and the bead surface is $1280 \pm 90~\mu\mathrm{m}$. For the small displacements studied here ($\le 850~\mu\mathrm{m}$) and given that $e<<D$, the deformation is close to a simple shear. The sample is illuminated by a laser sheet in the vertical $(x,y)$ plane, of thickness $\approx 100~\mu\mathrm{m}$. We image a region of size $710 \times 530~\mu\mathrm{m}^2$ using light scattered at $\theta = 90~\mathrm{deg}$, corresponding to $q = 20.6~\mu\mathrm{m}^{-1}$. To obtain space-resolved information on the mesoscopic displacement and the microscopic dynamics, we run our algorithm on ten ROIs of size $31 \times 264~\mu\mathrm{m}^2$ regularly spaced at a growing distance $x$ from the moving wall.

Figure~\ref{FIG:sheared}a shows the velocity profiles close to the bead wall, for two times $t$ after starting shearing the sample. In this representation, the slope of the data is the local shear rate $\dot{\gamma}$. The solid line shows the velocity profile expected for homogeneous shear, corresponding to an average shear rate across the whole gap of $\dot{\gamma} = 7.9  \times 10^{-5}~\mathrm{s}^{-1}$. It is clear that already at $t=120~\mathrm{s}$ $\dot{\gamma}$ is non-uniform across the gap, with a highly-sheared band close to the moving surface ($x \ge -164~\mu \mathrm{m}$, $\dot{\gamma} \approx 1.4  \times 10^{-4}~\mathrm{s}^{-1}$),  followed by a low shear region ($x \le -220 ~\mu \mathrm{m}$, $\dot{\gamma} \approx 6.0  \times 10^{-5}~\mathrm{s}^{-1}$). Similar shear banding has been reported for other colloidal glasses~\cite{chikkadi11}. Interestingly, shear banding is seen to evolve with time. At $t=820~\mathrm{s}$, the shear rate for the high- and low-shear bands is comparable to that at $t=120~\mathrm{s}$ ($\dot{\gamma} \approx 1.3 \times 10^{-4}~\mathrm{s}^{-1}$ and $\dot{\gamma} \approx 6.1  \times 10^{-5}~\mathrm{s}^{-1}$, respectively), but the boundary between the two zones has moved from $x = -190~\mu \mathrm{m}$ to $x = -240~\mu \mathrm{m}$. Additionally, the occurrence of a marked drop of $v_y$ close to the moving surface suggests slipping. This behavior is reminiscent of that reported for a variety of jammed or glassy soft materials, see \textit{e.g.}~\cite{divoux10}, which exhibit complex spatio-temporal shear patterns. Figure~\ref{FIG:sheared}b) shows the intensity correlation function measured for the ROI at the position indicated by the arrow in a). For the unsheared sample (crosses) no dynamics is observed on time scales up to $20 s$, about 2000 times the Brownian time for the same particles in the diluted regime. This is consistent with the notion that the microscopic dynamics of a sample at rest is slowed down by orders of magnitude on approaching the glass transition. The open symbols show the uncorrected $g_2-1$: a fast decay is observed, essentially due to the translation of the speckle pattern due to the imposed shear. Once corrected, the data still show a decay of $g_2-1$ (albeit a slower one), thus indicating that particles move with respect to each other, in addition to be advected by the shear deformation. We emphasize that the corrected $g_2-1$ is sensitive to the component of the particle displacement along the direction of $\mathbf{q}$, which lays in the horizontal plane, perpendicular to the shear direction. Therefore, the decay of $g_2-1$ is not due to the affine component of the particle displacement along $y$, but rather to irreversible rearrangements associated with flow in glassy systems~\cite{schall07}. Interestingly, we find that the decay of $g_2-1$ is faster at $t=820~\mathrm{s}$, when both the local $\dot{\gamma}$ and its gradient are larger. This suggest a direct relation between (local) shear rate and plastic rearrangements, as proposed for granular materials~\cite{amon12} and emulsions~\cite{jop12}.

\section{Conclusions}
We have introduced a method to obtain the mesoscopic displacement field and the microscopic dynamics in soft materials where the constituents undergo both a drift motion and a relative displacement. The algorithm proposed here is optimized for the typical features of speckle images in PCI experiments, where a small speckle size is highly desirable to maximize the spatial resolution and the statistics of the measurement. The algorithm is highly efficient in that the correction of $g_2-1$ does not requires any significant computational extracost, besides that necessary to determine the displacement field. The method has been successfully tested on a Brownian suspension and a colloidal glass. Although similar information may be in principle obtained using confocal or optical microscopy, our method allows one to investigate samples that are difficult or impossible to visualize in real space, such as the very small particles of our colloidal glass. A generalization to speckle patterns obtained under partially coherent illumination, such as in recent microscopy developments~\cite{cerbino09,giavazzi09} is also possible~\cite{Buzzaccaro}. The method presented here should be particularly valuable for soft materials where slow dynamics is coupled to the effects of an external stress, as in rheological experiments or in samples submitted to an external field such as gravity~\cite{brambilla11}, or in disordered jammed materials, where internal stress is known to play a major role~\cite{lieleg11,BouchaudEPJE2001} in the sample dynamics.

\section{Acknowledgements}
Funding from CNES is gratefully acknowledged. We thank Unilever for partially supporting SM.

\section{APPENDIX A: Center-of-mass algorithm for determining the direction of
shift}
\label{sec:centerofmass}

In our speckle images the speckle size is comparable to the pixel
size. Hence, the peak of $\crJI$ extends over a few pixels at most.
Accordingly, we calculate the center of mass of the peak based on
the values of $\crJI$ at its maximum, located at $(\km,\lm)$, and in
the eight neighboring pixels as showed in
Fig.~\ref{FIG:imageshift}b. Our aim is to determine in which of the
four quadrants A, B, C, and D shown in Fig.~\ref{FIG:imageshift}b
lays the center of mass of the correlation peak. To avoid any bias
introduced by the square shape and the orientation of the pixels, we adopt a circular
symmetry by considering only the contribution of the areas indicated by $\alpha_1, \alpha_2, \alpha_3$ in
Fig.~\ref{FIG:imageshift}b (for clarity, only the overlap areas for quadrant B are shown in the figure). This is accomplished by weighting the contribution of each element of $\crJI$ by its overlap with the
circle shown in the figure. The weights $w_A$, $w_B$, $w_C$, $w_D$
associated with quadrants A, B, C, D respectively are then
\begin{eqnarray}
w_A = \alpha_2 \crJI(\km-1,\lm-1) + \alpha_1[\crJI(\km-1,\lm) + \crJI(\km,\lm-1) + \alpha_3 \crJI(\km,\lm) \label{Eq:wA}\\
w_B = \alpha_2 \crJI(\km-1,\lm+1) + \alpha_1[\crJI(\km-1,\lm) + \crJI(\km,\lm+1) + \alpha_3 \crJI(\km,\lm) \label{Eq:wB}\\
w_C = \alpha_2 \crJI(\km+1,\lm-1) + \alpha_1[\crJI(\km,\lm-1) + \crJI(\km+1,\lm) + \alpha_3 \crJI(\km,\lm) \label{Eq:wC}\\
w_D = \alpha_2 \crJI(\km+1,\lm+1) + \alpha_1[\crJI(\km,\lm+1) +
\crJI(\km+1,\lm) + \alpha_3 \crJI(\km,\lm) \label{Eq:wD}\,,
\end{eqnarray}
with $\alpha_1 = 0.485869913$, $\alpha_2 = 0.545406041$,
$\alpha_3=0.25$. Once the weights
of the four quadrants are determined, the indexes to be used in
Eq.~(\ref{Eq:chi22}) and following are calculated from
\begin{eqnarray}
k_1 = k_2 = \mathrm{floor}(\km+\delta r)\\ \label{Eq:k1}
k_3 = k_4 = k_1+1\\
l_1 = l_3 = \mathrm{floor}(\lm+\delta c)\\ \label{Eq:l1}
l_2 = l_4 = l_1+1 \,,
\end{eqnarray}
where $\delta r $ and $\delta c$ are obtained from $w_A,...,w_D$:
\begin{eqnarray}
\delta r = \frac{w_C+w_D-w_A-w_B}{w_A+w_B+w_C+w_D}\\
\delta c = \frac{w_B+w_D-w_A-w_C}{w_A+w_B+w_C+w_D} \,.
\end{eqnarray}

\end{document}